\documentstyle[12pt]{article}
\def\be{\begin{equation}}
\def\ee{\end{equation}}
\begin{document}
\title{
\begin{flushright}
{\small SMI-09-98 }
\end{flushright}
\vspace{2cm}
On the Breaking of Conformal Symmetry in\\ the AdS/CFT Correspondence }
\author{ I.Ya.
Aref'eva  and I.V. Volovich\\ $~~~~$  \\ {\it Steklov Mathematical
Institute, Russian Academy of Sciences}\\ {\it Gubkin St.8, GSP-1,
117966,
Moscow, Russia}\\ arefeva, volovich@mi.ras.ru} \date {$~$} \maketitle
\begin {abstract}
The renormalization of the boundary action
in the AdS/CFT correspondence is considered and the breaking
of conformal symmetry is discussed.
\end {abstract}

\newpage
\section {Introduction}

A correspondence between the large $N$ limit of
certain superconformal theories and supergravity theories
on the boundary of Anti-de Sitter spaces has been conjectured
in \cite {Mal}. An  AdS/CFT
correspondence was used in
\cite{SfSk} in context of
physics of non-extremal black holes.

An  AdS/CFT relationship between correlation functions
in the large $N$ CFT and the classical supergravity
action on the boundary was proposed in
\cite{GKP,Wit1} in terms of the component supergravity and
in \cite{FeFrZa} in superspace formulation.
The proposal is that the generating functional
of the correlation functions in the large $N$ CFT is given
by  the supergravity
action considering as a functional of the boundary
values of the fields.

Construction of
the $n$-point correlation functions for scalar
fields within the proposal \cite{Wit1} has been  presented in
\cite{AV}  and it is considered in more detail in
\cite{MV}. The 3-point amplitudes including vector
fields  in the AdS supergravity are considered in \cite{FrMaMa}. The
free spinor fields are discussed in \cite{HeSf} and amplitudes
including graviton are considered in \cite{Tse}.
For other recent considerations of the AdS/CFT correspondence see
\cite{IMSY}-\cite{DaPo}.

To make the proposal more precise one
has to explain the meaning of the boundary action.
The boundary action  is not well defined for free fields
and one gets divergencies when one considers
the value of the classical action on the solutions
of the corresponding boundary problem.  In \cite{AV}
it was proposed to make a renormalization of the boundary
action to get the well defined functional and the breaking
of conformal invariance has been pointed out. In \cite {AV}
only massless scalar field have been considered.
In this note we shall discuss
the renormalization of the boundary action for
the massive scalar fields.

In \cite{GuHaKl} the breaking of conformal invariance by the irrelevant
Born-Infeld corrections to the Yang-Mills theory is considered. These
effects go beyond ``the strict conformal limit''. In this note
we discuss the breaking of conformal invariance that takes place
already in the strict conformal limit.

 Actually the proposals for the AdS/CFT correspondence
in  \cite{GKP} and
in \cite {Wit1} are somewhat different. In \cite{GKP}
a cut-off on the boundary of AdS is introduced
and to get the 2-point correlation function in CFT
it is proposed to take only ``the leading non-analytical
term'' in the Fourier transform of the boundary
action. Unfortunately this ``non-analytic'' prescription
depends on the used regularisation procedure and more specifically
on the cut-off.

In \cite {Wit1} an elegant expression for the
boundary action for free fields has
been proposed.
In the four dimensional Euclidean space for the massless fields
it has the form
\be
\label{1}
I(\Phi_0)=\frac{C}{2}\int \frac{\Phi_0({\bf x})\Phi_0({\bf y})}
{|{\bf x}-{\bf y}|^8}d{\bf x}d{\bf y}
\ee
Here $\Phi_0({\bf x})$ is a test function in the four dimensional space
which is considered as the boundary function for a field in the AdS
space.

The integral (\ref {1}) requires a regularisation. If one
interprets it as the value of the distribution $|{\bf x}|^{-8}$
on the test function then one has  the breaking of
the conformal symmetry because the distribution $|{\bf x}|^{-8}$
is not a homogeneous generalized function, see \cite{AV}
for a discussion of this point. Let us recall that the function
$|{\bf x}|^{-8}$ of course is scaling invariant for ${\bf x}\neq 0$ but
in (\ref{1}) one has the integral and it leads to divergencies
if we don't regularise it. The regularisation leads to
the breaking of the conformal invariance.

In other words
if in CFT one has the correlator of the form
\be
\label{2}
<{\cal O}({\bf x}){\cal O}({\bf y})>=\frac{C}{|{\bf x}-{\bf y}|^8}
\ee
then it is scaling invariant for ${\bf x}\neq {\bf y}$ because one
can consider it as an ordinary function. However
if we want to define the correlation function
 also for ${\bf x}={\bf y}$
in the distribution sense (for example if we want
to integrate it as in (\ref{1})) then one has the breaking
of the conformal symmetry.

In \cite {AV} the following renormalized boundary action  is
obtained (see also the discussion in the next section)
\be
\label{3}
I_{ren}(\Phi_0)=    \int _{R^4}d{\bf x}
 \Phi _0({\bf x})[a\Delta^2\log (-\Delta) +b\Delta^2 +c\Delta]
\Phi _0({\bf x})
\ee
where $\Delta$ is the Laplacian and
$a,b,c$ are arbitrary constants.

The action (\ref{3}) is not conformally invariant. If we
assume 'the minimal breaking of conformal symmetry''
(or  a minimal subtraction
renormalization scheme \cite {AV}) and take $c=0$
then we obtain the renormalized boundary action
\be
\label{4}
I_{ren}(\Phi_0)=    \int _{R^4}d{\bf x}
 \Phi _0({\bf x})[a\Delta^2\log (-\Delta)+b\Delta^2]
\Phi _0({\bf x})
\ee
The renormalized action (\ref{4}) (or (\ref{3}))  gives an
interpretation
of the formula (\ref{1}) if we  treat $|{\bf x}-{\bf y}|^{-8}$
as a distribution because the Fourier transform of the distribution
  $|{\bf x}|^{-8}$ is $ap^4 \log p +bp^4$.

\section {Massless Fields}

Let us recall first the renormalization of the boundary action
for massless fields \cite{AV}. The AdS action for this theory
in the Lobachevsky space $ R^{d+1}_+=\{(x_0, {\bf x}) \in R^{d+1}|x_0
>0\}$
is given by
\be
\label{Ads}
I=\frac{1}{2}\int _{\epsilon}^{\infty}dx_0
\int _{R^d}d{\bf x} \frac{1}{x_{0}^{d-1}}\sum_{i=0}^{d}(\frac{\partial
\Phi}{\partial x_{i}})^{2}
\ee
Here $\epsilon >0$ is a cut-off, see
\cite{GKP}. The solution of the
Dirichlet problem
\be
\label{LDp}
(\sum _{i=0}^{d}\frac{\partial^2}{\partial x_{i}^{2}} -
\frac{(d-1)}{x_{0}}\frac{\partial}{\partial x_{0}}) \Phi=0,~~~
\Phi |_{x_0=0}=\Phi _{0}({\bf x})
\ee
can be represented
in the form
\be
\label{CD}
\Phi (x_0,{\bf x})=
cx_{0}^{d/2}\int _{R^d}d{\bf p}e^{i{\bf px}} |{\bf p}|^{\frac{d}{2}}
K_{\frac{d}{2}}(|{\bf p}|x_{0}){\tilde \Phi}_0 ({\bf p}),
\ee
where $K_{\frac{d}{2}}(y)$ is the
modified Bessel function.
By integrating by parts, one can rewrite
(\ref{Ads}) as

\be
\label{Adsa1}
I=-\frac{1}{2}\int _{R^d}  d{\bf x}
(\frac{1}{x_{0}^{d-1}}\Phi \frac{\partial \Phi}{\partial x_0})
|_{x_0=\epsilon}
\ee
Using the asymptotic expansion  of the modified Bessel function
one gets a regularised expression for the action.

For $d=4$ one has  for $x_{0}\to 0$

\be
\label{CDA}
\Phi (x_0,{\bf x})=
C\int _{R^4}d{\bf p}e^{i{\bf px}}
[2-\frac{1}{2}(x_0p)^2-\frac{(x_0p)^4}{8}\log
\frac{x_0p}{2}+c(x_0p)^4+~...~]
{\tilde \Phi}_0({\bf p}),
\ee
here $p= |{\bf p}|$.
The action (\ref{Adsa1}) for $\epsilon \to 0$ behaves as

\be
\label{bac}
I=
C\int _{R^4}d{\bf p}|{\tilde \Phi}_0({\bf p})|^2
[-\frac{1}{\epsilon ^2}p^2-\frac{p^4}{2}
\log\frac{\epsilon p}{2}+c_1p^4+~...~].
\ee

 The appearance of divergent
terms in the classical action  can be  related with
the fact that the propagator for a field ${\cal O}$
of conformal dimension 4
should be a multiple of $|{\bf x}-{\bf y}|^{-8}$ and
one has to define it as a distribution.

In the spirit of the minimal subtractions scheme in the  theory
of renormalization one can write a "renormalized"
action as

\be
\label{rac}
I_{ren}=
\int _{R^4}d{\bf p}|{\tilde \Phi}_0({\bf p})|^2
[ap^4 \log p^2+bp^4].
\ee

One can write the final result as follows
\be
\label{c}
I=\int _{R^5_+} dx\sqrt{g}(\nabla \Phi )^2
\longrightarrow
I_{ren}=    \int _{R^4}d{\bf x}
 \Phi _0[a\Delta^2\log (-\Delta)+b\Delta^2] \Phi _0
\ee
where the arrow includes the renormalization.
The renormalized action includes a local term
\be
\label{fac}
\int_{R^4}d{\bf x}(\Delta \Phi_0)^2
\ee
There is also a non-local term. This is related with the fact
that the distribution
$|{\bf x}|^{-8}$ \cite{GS} is not a homogeneous in $R^4$
(there is the logariphm in the scaling law).

If one adds also  finite parts, then one gets  a term
$\Phi _0\Delta \Phi _0$.
One requires additional physical
assumptions  to fix the form of the renormalized action.
There is an analogy with the choice of "the leading non-analitical term"
in \cite{GKP}.

One can consider the renormalized
action  (\ref{rac}) as a definition of distribution
$|\bf{x}-\bf{y}|^{-8}$
and  interpret the action \cite{Wit1}
\be
\label{wa}
\int_{R^{2d}} \frac{\Phi _0({\bf x})\Phi _0({\bf y})}
{|{\bf x}-{\bf y}|^{2d}} d{\bf x}d{\bf y}
\ee
as the value of the distribution $|{\bf x}-{\bf y}|^{-2d}$
on a test function.
The Fourier transform of this distribution \cite{GS}
includes a logariphmic term. For $d=4$  one has
\be
\label{ftr}
\widetilde {|{\bf x }|^{-8}} =ap^4\log p +bp^4,
\ee

The distribution $|{\bf x}|^{-8}=r^{-8}$ is the so called
associated generalized function. It is obtained by taking the
analytical continuation over $\lambda$ of the distribution
$r^{\lambda}$ and then expanding at a pole. In the $d$-dimensional space
one has at $\lambda=-d-2k$ the expansion
\be
r^{\lambda}=\omega_d [\frac{\delta^{(2k)}(r)}{(2k)!}\frac{1}
{\lambda+d+2k}+r^{-d-2k}+(\lambda+d+2k)r^{-d-2k}\log r+...]
\ee
where $\omega_d=2\pi^{d/2}/\Gamma (d/2)$.

\section {Massive Fields}

Let us discuss now the  massive case. The  free action in the AdS space
has the form
\be
\label{Adsm}
I=\frac{1}{2}\int _{\epsilon}^{\infty}dx_0
\int _{R^d}d{\bf x}[ \frac{1}{x_{0}^{d-1}}\sum_{i=0}^{d}(\frac{\partial
\phi}{\partial x_{i}})^{2} + \frac{m^2}{x_{0}^{d+1}}\phi ^2 ]
\ee
Let us consider a solution of the following boundary problem
\be
\label{LDpm}
(\sum _{i=0}^{d}\frac{\partial^2}{\partial x_{i}^{2}} -
\frac{(d-1)}{x_{0}}\frac{\partial}{\partial x_{0}}
-\frac{m^2}{x_{0}^2}) \Phi=0,~~~
(x_0)^{k}\Phi |_{x_0=0}=\Phi _{0}({\bf x})
\ee
where $k$ is the larger root of the equation
\be
m^2=k(d+k),
\ee
This solution can be represented in the form
\be
\label{CDm}
\Phi (x_0,{\bf x})=
cx_{0}^{d/2}\int _{R^d}d{\bf p}e^{i{\bf px}} |{\bf p}|^{\frac{d}{2}+k}
K_{\Delta}(|{\bf p}|x_{0}){\tilde \Phi}_0 ({\bf p}),
\ee
where
\be
\label{mass}
\Delta=\sqrt{(\frac{d}{2})^2+m^2},~~~~~
\ee
Let us consider the case of integer $k$.
For even $d$ by using the expansion for $x_0\to 0$

$$
K_{\frac{d}{2}+k}(x_0p)=
\frac{1}{(x_0p)^{\frac{d}{2}+k}}[\alpha _0+\alpha _1x_0p
+...]+
$$
\be
\log (x_0p)\cdot (x_0p)^{\frac{d}{2}+k}
[\beta_0+\beta _1 x_0p+...]+
(x_0p)^{\frac{d}{2}+k}[\gamma _0+\gamma _1 x_0p+...]
\ee
we get
$$
\Phi (x_0,{\bf x})=
c\int _{R^d}d{\bf p}e^{i{\bf px}} [\frac{1}{(x_0
p)^{k}} (\alpha_0 +\alpha _1x_0p+...)
$$
\be
\label{CDma}
+(x_0p)^{d+k}\log(x_0p)(\beta _0+\beta _1 (x_0
p+...)+x_0p(\gamma_0+\gamma_1 x_0p+...)]
p^k{\tilde
\Phi}_0 ({\bf p}),
\ee
and
\be
(x_0)^k\Phi (x_0,{\bf x})|_{x_{0}=0}=
c\alpha_0\int _{R^d}d{\bf p}e^{i{\bf px}}{\tilde \Phi}_0 ({\bf p})
\ee

By integrating
by parts, one can rewrite (\ref{Adsm}) as

\be
\label{Adsa}
I=-\frac{1}{2}\int _{R^d}  d{\bf x}
(\frac{1}{x_{0}^{d-1}}\Phi \frac{\partial \Phi}{\partial x_0})
|_{x_0=\epsilon}
\ee
Using the asymptotic expansion  of the modified Bessel function
one gets a regularised expression for the action.
By using (\ref{CDma}) we obtain that
the action (\ref{Adsa}) for $\epsilon \to 0$ behaves as

\be
\label{bacm}
I=
\int _{R^4}d{\bf p}|(p)^{2k}{\tilde \Phi}_0({\bf p})|^2
[\frac{1}{\epsilon ^{d}}\sum _{n=-2k}^{d}
(p\epsilon )^{n}a_n+
ap^{d}\log(\epsilon p)].
\ee

Performing  the minimal subtractions as in the massless case
we get the renormalized boundary action

\be
\label{racm}
I_{ren}=
\int _{R^d}d{\bf p}|{\tilde \Phi}_0({\bf p})|^2
[ap^{d+2k} \log p+bp^{d+2k}].
\ee

In principle we have to add to (\ref{racm}) also other terms
corresponding to the divergent terms in (\ref{bac}).

One can write the final result as follows
\be
\label{cm}
I=\int _{R^{d+1}_+} dx\sqrt{g}[(\nabla \Phi )^2+m^2( \Phi )^2]
\longrightarrow
I_{ren}=    \int _{R^d}d{\bf x}
 \Phi _0(-\Delta )^{d/2+k}[a\log(-\Delta)+b] \Phi _0
\ee
where the arrow includes the renormalization.
If one adds also  finite parts, then one gets   terms
$\Phi _0(-\Delta )^{l}\Phi _0$, $l=1,....,d/2+k-1$.
One requires additional physical
assumptions  to fix the form of the renormalized action.
As for massless case the renormalized action includes  the local term
as well as  the non-local term. The origine of this fact  is
that the distribution $|{\bf x}|^{-2d-2k}$ \cite{GS} is not a
homogeneous in $R^d$ (there is the logariphm in the scaling law).

One can consider the renormalized
action  (\ref{rac}) as a definition of distribution
$|\bf{x}-\bf{y}|^{-2(d+k)}$ and  interpret the action \cite{Wit1}
\be
\label{wam}
\int_{R^{2d}} \frac{\Phi _0({\bf x})\Phi _0({\bf y})}
{|{\bf x}-{\bf y}|^{2d+2k}} d{\bf x}d{\bf y}
\ee
as the value of the distribution $|{\bf x}-{\bf y}|^{-2d-2k}$
on a test function.
The Fourier transform of this distribution \cite{GS}
includes a logariphmic term. For $d$ even one has
\be
\label{ftrm}
\widetilde {|{\bf x }|^{-2(d+k)}} =ap^{d+2k}\log p+bp^{d+2k}.
\ee

\section {Conclusion}

In this note the renormalized boundary action (\ref{racm}),(\ref{cm})
for the free massive
scalar field is obtained.  There is the breaking of the conformal
symmetry in the AdS/CFT correspondence of the same type as
it was found earlier for the massless case \cite{AV}.
In this note only the simplest examples of the AdS/CFT correspondence
have been  discussed. It is interesting to study how the
consideration of interaction and the $n$-point correlation functions
( let us recall that in the standard background formalism
\cite{AFS} the behaviour of the
two-point function is more subtle then the $n$-point functions)
and also supersymmetry will affect the breaking of  conformal symmetry.

The work is supported in part by
INTAS grant 96-0698. I.A. is supported also by RFFI grant 96-01-00608
and I.V. is supported  in part by  RFFI grant 96-01-00312.


{\small
\begin{thebibliography}{99}
\bibitem{Mal} J. Maldacena, The large {N} limit of superconformal field
theories and  supergravity, hep-th/9711200

\bibitem{SfSk}  K. Sfetsos and K. Skenderis,
Microscopic derivation of the Bekenstein-Hawking entropy
formula for non-extremal black holes, hep-th/9711138,

H.J. Boonstra, B. Peeters and K. Skenderis, hep-th/9801076

\bibitem{GKP}
S.S. Gubser, I.R. Klebanov and A.M. Polyakov, Gauge theory correlators
from noncritical string theory, hep-th/9802109

\bibitem{Wit1} E.Witten, The large N limit of superconformal field
theories and supergravity, hep-th/9802150

\bibitem{FeFrZa} S.Ferrara, C. Fronsdal and A. Zaffaroni,
On N=8 Supergravity On Ads(5) And N=4 Superconformal Yang-Mills
Theory, hep-th/9802203.

\bibitem{AV} I.Ya. Aref'eva and I.V. Volovich,
On Large N Conformal Theories, Field Theories In Anti-De Sitter
Space And Singletons, hep-th/9803028

\bibitem{MV} W. M\" uck and K.S. Viswanathan,
Conformal field theory correlators from classical scalar field
theory on AdS$_{d+1}$, hep-th/9804035.

\bibitem{HeSf}  M.Henningson and K.Sfetsos,
Spinors and the {AdS/CFT} correspondence, hep-th/9803251.

\bibitem{FrMaMa} D.Z. Freedman, S.D. Mathur, A. Matusis and L. Rastelli,
Correlation functions in the CFT$_d$/AdS$_{d+1}$
correspondence, hep-th/9804058.

\bibitem{Tse} Hong Liu and  A.A. Tseytlin, D=4 Super Yang-Mills,
D=5 gauged supergravity,
and $D=4$ conformal supergravity, hep-th/9804083

\bibitem{GS} I.M.Gelf'and and G.E.Shilov, Generalized Functions,
Academic press, 1963

\bibitem {IMSY} N. Itzhaki, J. M. Maldacena, J. Sonnenschein, and
S. Yankielowicz,  hep-th/9802042.

\bibitem{GuMi} M. Gunaydin and D. Minic,
Singletons, Doubletons and M-theory, hep-th/9802047

\bibitem{HoOo} G. T. Horowitz and H. Ooguri,
Spectrum of Large N Gauge Theory from Supergravity, hep-th/9802116

\bibitem{FrFe} S. Ferrara and  C. Fronsdal,
Gauge fields as composite boundary
excitations, hep-th/9802126.

\bibitem{KaSi} S. Kachru, E. Silverstein,
`4-D Conformal Theories And Strings On Orbifolds', hep-th/9802183

\bibitem{Ber} M. Berkooz,
`A Supergravity Dual Of A (1,0) Field Theory In Six-Dimensions',
hep-th/9802195

\bibitem{BL} V. Balasubramanian and F. Larsen, hep-th/9802198

\bibitem{Rey} S.-J. Rey, J. Yee,
`Macroscopic Strings As Heavy Quarks In Large N Gauge Theory And
Anti-De Sitter Supergravity', hep-th/9803001

\bibitem{Mal2} J. Maldacena,
`Wilson Loops In Large N Field Theories', hep-th/9803002
\bibitem{FlFr} M. Flato, C. Fronsdal,
`Interacting Singletons', hep-th/9803013

\bibitem{LNV} A. Lawrence, N. Nekrasov and C. Vafa, hep-th/9803015

\bibitem{GuHaKl} S. Gubser, A. Hashimoto, I.R. Klebanov,
 M. Krasnitz,
`Scalar Absorption And The Breaking Of The World Volume
Conformal Invariance', hep-th/9803023

\bibitem{CaCeAu} L. Castellani, A. Ceresole, R. D'Auria,
S. Ferrara, P. Fre, M. Trigiante, `G/H M-Branes And Ads(P+2)
Geometries', hep-th/9803039

\bibitem{AOY} Ofer Aharony, Yaron Oz, Zheng Yin,
`M Theory On $AdS_p\times S^{11-p}$ and Superconformal Field Theories',
hep-th/9803051

\bibitem{Min} S. Minwalla,
`Particles On Ads(4/7) And Primary Operators On M(2)-Brane And
M(5)-Brane World Volumes',  hep-th/9803053

 \bibitem{FeZa} S. Ferrara, A. Zaffaroni,
N=1,2 4D Superconformal Field Theories and Supergravity in $AdS_5$
 hep-th/9803060

\bibitem{Leigh}
R.G. Leigh and M.Rozali, The large N limit of the (2,0)
superconformal field theory. hep-th/9803068

\bibitem{BKV} M. Bershadsky, Z. Kakushadze, C. Vafa,
String Expansion as Large N Expansion of Gauge Theories,
hep-th/9803076.

\bibitem{Ha} E. Halyo, Supergravity
on $AdS_{4/7} \times S^{7/4}$ and M Branes, hep-th/9803077

\bibitem{Raj} A. Rajaraman,
Two-Form Fields and the Gauge Theory Description of Black Holes,
hep-th/9803082

 \bibitem{BeBe} E. Bergshoeff, K. Behrndt, D-Instantons and asymptotic
geometries,
hep-th/9803090

\bibitem{FKPZ} S. Ferrara, A. Kehagias, H. Partouche, A. Zaffaroni,
 Membranes and Fivebranes with Lower Supersymmetry
and their AdS Supergravity Duals, hep-th/9803109

\bibitem{Go} J. Gomis, Anti de Sitter
Geometry and Strongly Coupled Gauge Theories, hep-th/9803119

\bibitem{Wit2} E. Witten,
Anti-de Sitter Space, Thermal Phase Transition, And Confinement In Gauge
Theories, hep-th/9803131

\bibitem{RTY} S.J.Rey. S.Theisen and J.T.Yee, 
Wilson-Polyakov Loop at Finite Temparature in Large N Gauge 
Theory and Anti-de-Sitter Supergravity,
hep-th/9803135.

\bibitem{BISY} A. Brandhuber, N.Itzhaki, J. Sonnenschein,
S. Yankielowicz, Wilson Loops, Confinement, and Phase Transitions in
Large N Gauge Theories from
Supergravity, hep-th/9803137


\bibitem{Gu} M. Gunaydin,
Unitary Supermultiplets of OSp(1/32,R) and M-theory,
    hep-th/9803138

\bibitem{OzTe} Y. Oz, J.Terning,
 Orbifolds of $ AdS_5xS^5$ and 4d Conformal Field Theories,
hep-th/9803167

 \bibitem{Igor} I.V. Volovich,
Large N Gauge Theories and Anti-de Sitter Bag Model, hep-th/9803174

\bibitem{Kak} Z. Kakushadze, Gauge Theories from Orientifolds and
Large N Limit,
hep-th/9803214.

\bibitem{Nastya}  A. Volovich,
 Near Anti-de Sitter Geometry and Corrections to the Large N Wilson Loop
 hep-th/9803220

\bibitem{BPS} H. Boonstra, B. Peeters, K. Skenderis,
Brane intersections, anti-de Sitter spacetimes and dual
superconformal theories, hep-th/9803231

\bibitem{FKPZS} S. Ferrara, A. Kehagias, H. Partouche, A. Zaffaroni,
$AdS_6$ Interpretation of 5d Superconformal Field Theories,
hep-th/9804006

\bibitem{MS} J. Maldacena, A. Strominger
 AdS3 Black Holes and a Stringy Exclusion Principle,
hep-th/9804085

\bibitem{Iv} V. D. Ivashchuk,
Composite p-branes on Product of Einstein Spaces, hep-th/9804113

\bibitem{DaPo} Ulf H. Danielsson, A. P. Polychronakos,
 Quarks, monopoles and dyons at large N, hep-th/9804141

\bibitem{AFS}I.Ya. Aref'eva, L.D. Faddeev and A.A. Slavnov,
Theor. Math. Phys.,21(1975)165
\end{thebibliography}
}
\end{document}